\documentclass[pra,showpacs,aps,amssymb,amsmath,nofootinbib,notitlepage,superscriptaddress,twocolumn,longbibliography]{revtex4-2}

\usepackage{graphicx,graphics,epsfig,times,bm,mathrsfs}
\usepackage[normalem]{ulem}
\usepackage{subcaption}
\usepackage{xfrac}
\usepackage{mathtools}
\usepackage{amsmath,amssymb,amsfonts}
\usepackage{varwidth}
\usepackage{babel}
\usepackage{hhline}
\usepackage{multirow}
\usepackage{pgf}
\usepackage{physics}
\usepackage{bbm}
\usepackage[hidelinks]{hyperref}
\usepackage{float}
\usepackage{microtype}
\usepackage[T1]{fontenc}
\usepackage[utf8]{inputenc}
\usepackage{algorithm}
\usepackage[noend]{algpseudocode}
\usepackage{cancel}

\usepackage{lineno}

\begin{document}

\title{Study of long-fiber Sagnac interferometers for twin field quantum key distribution networks}

\author{Reem Mandil}
\email{reem.mandil@mail.utoronto.ca} 
\affiliation{Centre for Quantum Information and Quantum Control, Department of Physics, University of Toronto, Toronto, Ontario, M5S 1A7, Canada}

\author{Li Qian}
\affiliation{Centre for Quantum Information and Quantum Control, Department of Electrical and Computer Engineering, University of Toronto, Toronto, Ontario, M5S 3G4, Canada}

\author{Hoi-Kwong Lo}
\affiliation{Centre for Quantum Information and Quantum Control, Department of Physics, University of Toronto, Toronto, Ontario, M5S 1A7, Canada}
\affiliation{Centre for Quantum Information and Quantum Control, Department of Electrical and Computer Engineering, University of Toronto, Toronto, Ontario, M5S 3G4, Canada}
\affiliation{Quantum Bridge Technologies, Inc., 100 College Street, Toronto, Ontario, M5G 1L5, Canada}

\date{\today}

\begin{abstract}
A Sagnac loop structure can help overcome the major difficulty in the practical implementation of a twin field quantum key distribution (TFQKD) network, namely, the need to stabilize the phase of a quantum state over many kilometers of fiber. Unfortunately, Rayleigh backscattering noise limits the signal-to-noise ratio for Sagnac systems containing long fibers and lossy photonic devices. Here, we solve this problem by sending optical pulses in long on-off bursts and using time post-selection on measurements taken with free-run single-photon avalanche detectors. We also investigate the impact of the residual phase noise uncompensated by the Sagnac structure and find that the variance of the phase noise scales as loop length to the third power, verifying an existing calculation in the literature. We measure the interference visibility in Sagnac loops of varying length without active phase stabilization and achieve $>97\%$ visibility in 200~km ultra-low-loss fiber, which is, to our knowledge, the longest fiber Sagnac interferometer demonstrated. Our results suggest that a Sagnac system is feasible for long-distance TFQKD networks, an important step towards the practical implementation of metropolitan quantum networks. 
\end{abstract}

\maketitle

\section{Introduction}\label{sec:intro}
Based on carriers that cannot be copied or eavesdropped without notice to the communicating parties, quantum key distribution (QKD) allows remote users to establish shared encryption keys with information-theoretic security~\cite{bb84,e91,Scarani2009,Xu2020}. QKD networks are an important building block for large-scale quantum networks and have been studied extensively~\cite{Townsend1997,Elliot2005,Peev2009,Sasaki2011,Frohlich2013,Tang2016,Chen2021,Mandil2023}. However, their key rates are limited by the repeaterless bounds on the key rate scaling with channel loss~\cite{Pirandola2017,Takeoka2014}. A variant of QKD, called twin field QKD (TFQKD), has attracted much scientific attention because it can beat these bounds~\cite{Lucamarini2018}, thus offering a promising approach to long-distance QKD networks.  

In TFQKD, two remote users (conventionally called Alice and Bob) send coherent pulses encoded with an optical phase to an untrusted central node (conventionally called Charlie) who performs a single-photon interference measurement on the states~\cite{Curty2021}, which requires phase stability between the two interfering optical paths. In most TFQKD experiments, Alice's and Bob's laser sources are phase-locked to a reference laser over additional fiber channels, forming a large Mach-Zehnder interferometer~\cite{Fang2020, Pittaluga2021, Liu2021, Wang2022, Liu2023}. Recently, there have also been demonstrations without phase locking the two sources~\cite{Li2023,Zhou2023,Chen2024}. For example, Ref.~\cite{Li2023} achieves this with the use of strong reference pulses and a fast Fourier transform algorithm to reconcile the phase difference in post-processing. In Ref.~\cite{Chen2024}, a local optical frequency standard is employed at each source to establish an absolute reference, a technique also used in post-measurement pairing QKD~\cite{Ge2024}. In all of the aforementioned works, however, expensive superconducting nanowire single-photon detectors with precise timing resolution are used in order to measure~\cite{Fang2020,Liu2021,Liu2023,Li2023,Chen2024} and/or compensate~\cite{Pittaluga2021,Wang2022,Zhou2023} phase noise. Moreover, it is practically challenging to stabilize interferometers with highly asymmetric long-fiber links, a situation that is inevitable in a multi-user-pair TFQKD network, and this task has only been demonstrated using sub-Hz linewidth lasers with matched frequencies~\cite{Zhou2023}. 

Unlike the other systems, TFQKD based on a Sagnac interferometer shows promise for easy networking and low-cost implementation~\cite{Zhong2019, Zhong2021, Zhong2022PRA} due to common-path stable interference. In this configuration, a single laser source (possessed by Charlie) is used to distribute coherent pulses into a fiber ring where one user encodes the clockwise pulses and the other user encodes the counterclockwise pulses. However, Sagnac TFQKD with auto phase stabilization has been restricted to fiber distances of less than 20~km. There are two main challenges to implementing Sagnac TFQKD over long fibers. One is the uncompensated fiber phase noise due to temperature fluctuations and vibrations in the surrounding environment that cause the phase in the fiber to fluctuate during the loop transit time. The other is the Rayleigh backscattering noise in fiber, a challenge also encountered in other TFQKD systems~\cite{Chen2020,Wang2022} but which is especially problematic in a Sagnac loop due to the bidirectional pulse transmission. This noise is proportional to the intensity and repetition rate of the source at Charlie and thus, for a high-speed and long-distance Sagnac TFQKD network (which contains several lossy photonic devices), the SNR at the detectors becomes limited by the backscattering.

We note that Park \textit{et al.}~\cite{Park2022} demonstrated a TFQKD network based on a Sagnac configuration with 160~km fiber, where they employ a star topology as opposed to the ring topology of Ref.~\cite{Zhong2022PRA}. However, they report that the phase stability by the common-path nature was not observed in their experiment and therefore they used a phase post-compensation method to compensate for the fiber phase drift. A recent work by Bertaina~\textit{et al.}~\cite{Bertaina2023} investigated the phase noise in TFQKD systems such as those implemented in Refs.~\cite{Fang2020, Pittaluga2021, Liu2021, Wang2022, Liu2023,Li2023,Zhou2023}, but did not consider Sagnac-based TFQKD systems. Previously, Min\'a\ifmmode \check{r}\else \v{r}\fi{}~\textit{et al.} measured the phase noise and visibility in Sagnac interferometers with fiber lengths up to 72~km~\cite{Minar2008}. In their work, a pulsed source at low repetition rate (kHz) was used to ensure that the backscattering noise did not limit the SNR. Bogdanski~\textit{et al.} demonstrated BB84 QKD over a Sagnac loop length of 150~km, using an active polarization control system to compensate the fiber birefringence effects~\cite{BOGDANSKI2009}. In their setup, they limited the backscattering noise by using an optical attenuator to reduce the source intensity. Qi~\textit{et al.} also employed a Sagnac interferometer for BB84 QKD, with a fiber length of 40~km and a low pulse repetition rate (kHz)~\cite{Qi2006}. 

Indeed, no existing literature assesses the feasibility of implementing high-speed long-distance Sagnac TFQKD, with respect to auto phase stability as well as Rayleigh backscattering noise. We overcome both the phase noise and the Rayleigh backscattering noise in fiber to achieve stable interference in loop lengths reaching 200~km. We mitigate the fiber phase noise via thermal and vibrational isolation of the fiber spools; no active phase stabilization is performed, nor do we perform phase post-compensation. Polarization alignment is maintained without fast or automated control. We report measurements on the phase noise as a function of Sagnac loop length, $L$, and find the variance to scale roughly as $L^3$. We circumvent the fiber backscattering noise by sending pulses in long, on-off bursts and using post-selection on time-resolved measurements taken with free-run single-photon avalanche detectors (SPADs). We achieve $>99\%$ visibility in 150~km fiber, and $>97\%$ visibility in 200~km. Our results represent record performance in terms of Sagnac interference over long fibers, and suggest that the Sagnac loop is feasible as a platform for long-distance TFQKD networks.

\section{Sagnac Interferometry in Fiber}\label{sec:sagnac}
In general, interferometric stability over long fiber channels is experimentally challenging since the refractive index and physical length of a fiber change in the presence of temperature fluctuations and vibrations in the surrounding environment~\cite{Ashby2007}. In a Sagnac interferometer, the interference is between two light fields (generated by the same source) counterpropagating in the same fiber loop. A phase change in a given segment of the fiber can affect both fields equally, leading to a zero phase difference. Note, however, this requires that the phase in the fiber does not fluctuate during the loop transit time, because the two fields travel through a given segment of the fiber at different times. Disturbances, even slow ones, that cause the fiber birefringence to fluctuate~\cite{Barlow1983} will result in polarization misalignment between the interfering fields because the fiber birefringence differently changes the polarization of the clockwise and counterclockwise light~\cite{Mecozzi2011}.

\begin{figure} 
    \begin{center}
       \includegraphics[width=0.99\linewidth]{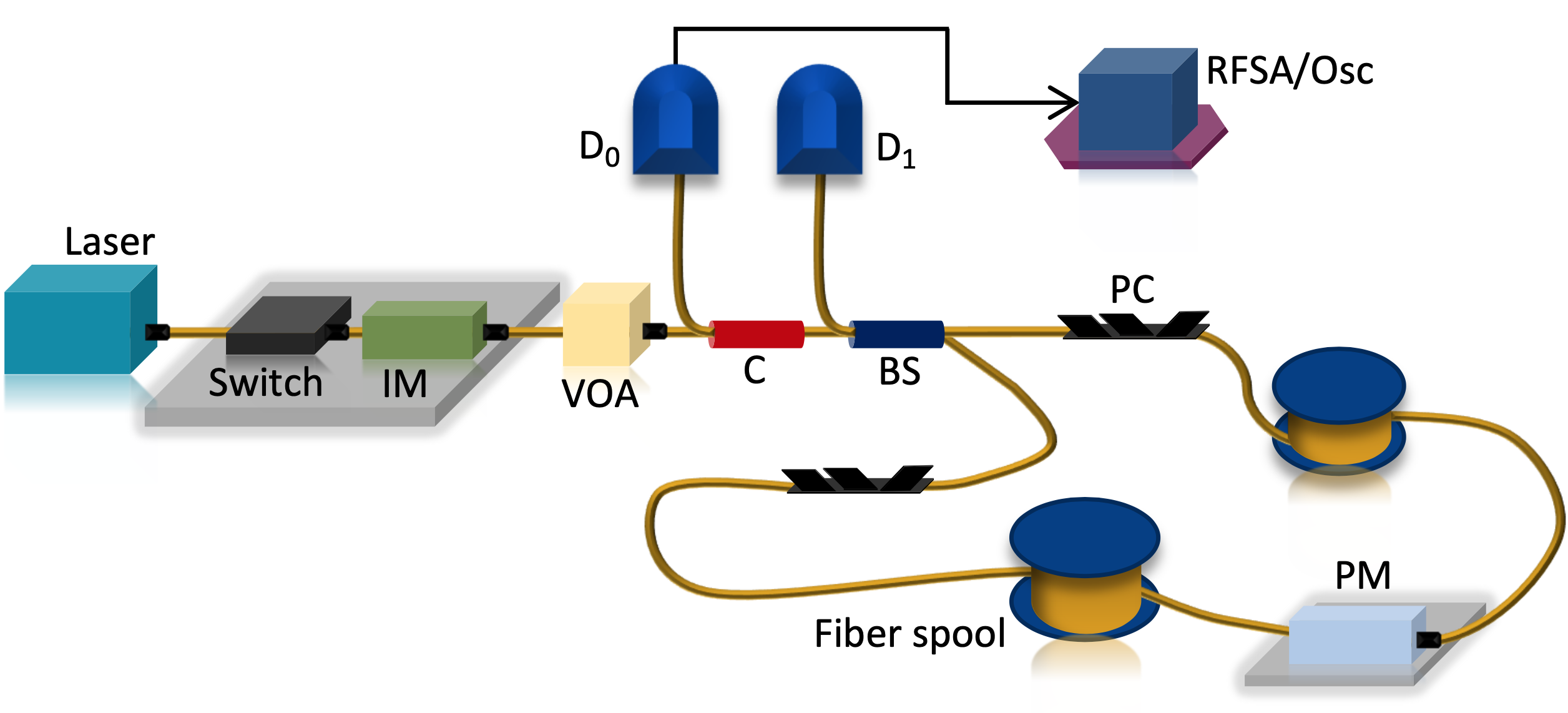} 
    \end{center}
    \caption{Sagnac interferometer experimental setup. Light from a continuous-wave laser launches into a fiber ring of length $L$ via a 50:50 beam splitter (BS). Interference of the counterpropagating paths is measured by detectors $D_0$ and $D_1$. Components on the hexagonal or square platforms are added when measuring phase noise or visibility, respectively. Classical photodetectors and single-photon avalanche detectors are used for phase noise and visibility measurements, respectively. IM: intensity modulator; VOA: variable optical attenuator; C: circulator; PC: polarization controller; PM: phase modulator; RFSA: rf spectrum analyzer; Osc: oscilloscope.} \label{fig:sagnac}
\end{figure}

Fig.~\ref{fig:sagnac} depicts a fiber Sagnac interferometer. Light from a continuous-wave (CW) laser is split at a 50:50 beam splitter and sent to a fiber ring. After traversing the loop, the counterpropagating fields interfere at the beam splitter. Polarization controllers (PCs) are adjusted such that both fields have the same polarization upon interference. The intensity at the constructive interference output of the interferometer is then given by~\cite{Minar2008}
\begin{equation}\label{eq:I_int}
    I(\delta \phi(t)) = (\frac{I_\text{max}-I_\text{min}}{2})\left[1+\cos\left(\phi + \delta \phi(t)\right) \right] + I_\text{min},
\end{equation} where $I_\text{max}$ and $I_\text{min}$ are the maximal and minimal measured intensity, respectively, $\phi$ is a constant phase difference between the two fields, and $\delta \phi$ is the phase fluctuation. Eq.~\ref{eq:I_int} is valid so long as perturbations to the fiber birefringence are small. This assumption is justified over time scales on the order of milliseconds, which applies to the phase fluctuation measurements presented in this work. 

Overall, fiber phase noise will cause the intensity of Sagnac interference $I(\delta \phi(t))$ to fluctuate as a function of time. The spectral density of these intensity variations is called the intensity noise power spectral density (PSD) and it can be measured with a RF spectrum analyzer (RFSA). As the transit time increases in longer fiber loops, more phase noise appears in the interference signal. We may also quantify the impact of phase noise by measuring the interference visibility,
\begin{equation}\label{eq:visibility}
    V = \frac{I_{\text{max}}-I_{\text{min}}}{I_{\text{max}}+I_{\text{min}}}.
\end{equation} As shown in Ref.~\cite{Minar2008}, the visibility is related to the variance of the phase fluctuation, $\sigma_{\delta\phi}^2$, by
\begin{equation}\label{eq:v_vs_var}
    V = e^{-\sigma_{\delta\phi}^2/2}.
\end{equation}

In addition to phase noise, the Sagnac interference signal will also be impacted by backscattering noise. Since this noise is phase-incoherent, it will make negligible contributions to the interference PSD. It will, however, reduce the interference visibility by raising the detector noise floor.

\section{Rayleigh Backscattering Noise}\label{sec:backscattering}
It is well known that Rayleigh scattering in the backward direction poses a significant challenge in two-way fiber optic QKD systems~\cite{Subacius2005}. For example, in a Sagnac loop, when clockwise pulses intersect with counterclockwise pulses, they will in general have very different intensities upon crossing. The weaker pulses (the ones that have travelled a greater length of the loop) will then be detected simultaneously with the backscattering of the stronger pulses. Since Rayleigh scattered light is of the same frequency as the incident light, it cannot be removed by means of optical filtering. Additionally, since the scattering can occur at any point along the fiber, it is not possible to gate it out of a detection window entirely. As a result, Rayleigh backscattering contributes to the false-click probability of the detectors. In order to solve this problem, we first develop a model for simulating the backscattering noise in a Sagnac interferometer using an experimentally-measured backscattering coefficient, then we adopt a practical technique that entails sending pulses in bursts. Details on the simulation and the measurement of the backscattering coefficient are found in Appendix~\ref{app:eta}.

To avoid backscattering noise rates that are higher than the signal rates, we may send pulses in bursts with a judiciously selected on-off time. This burst-patterning technique exploits the time-dependence of the backscattering noise to ensure that when the pulses arrive at the detectors, the backscattering noise has decayed to a tolerable level. Our backscattering simulation allows us to choose a burst pattern that will result in a high SNR for an arbitrary Sagnac system. We note that our solution to the Rayleigh backscattering problem in Sagnac TFQKD is reminiscent of the one proposed by Ribordy~\textit{et al.} for ``plug \& play'' QKD~\cite{Ribordy2000} and employed in Ref.~\cite{Park2022}, with a key difference to be highlighted in the Discussion. 

\section{Fiber Phase Noise}\label{sec:psd} 
\begin{figure} [htbp]
    \begin{center}
       \includegraphics[width=0.99\linewidth]{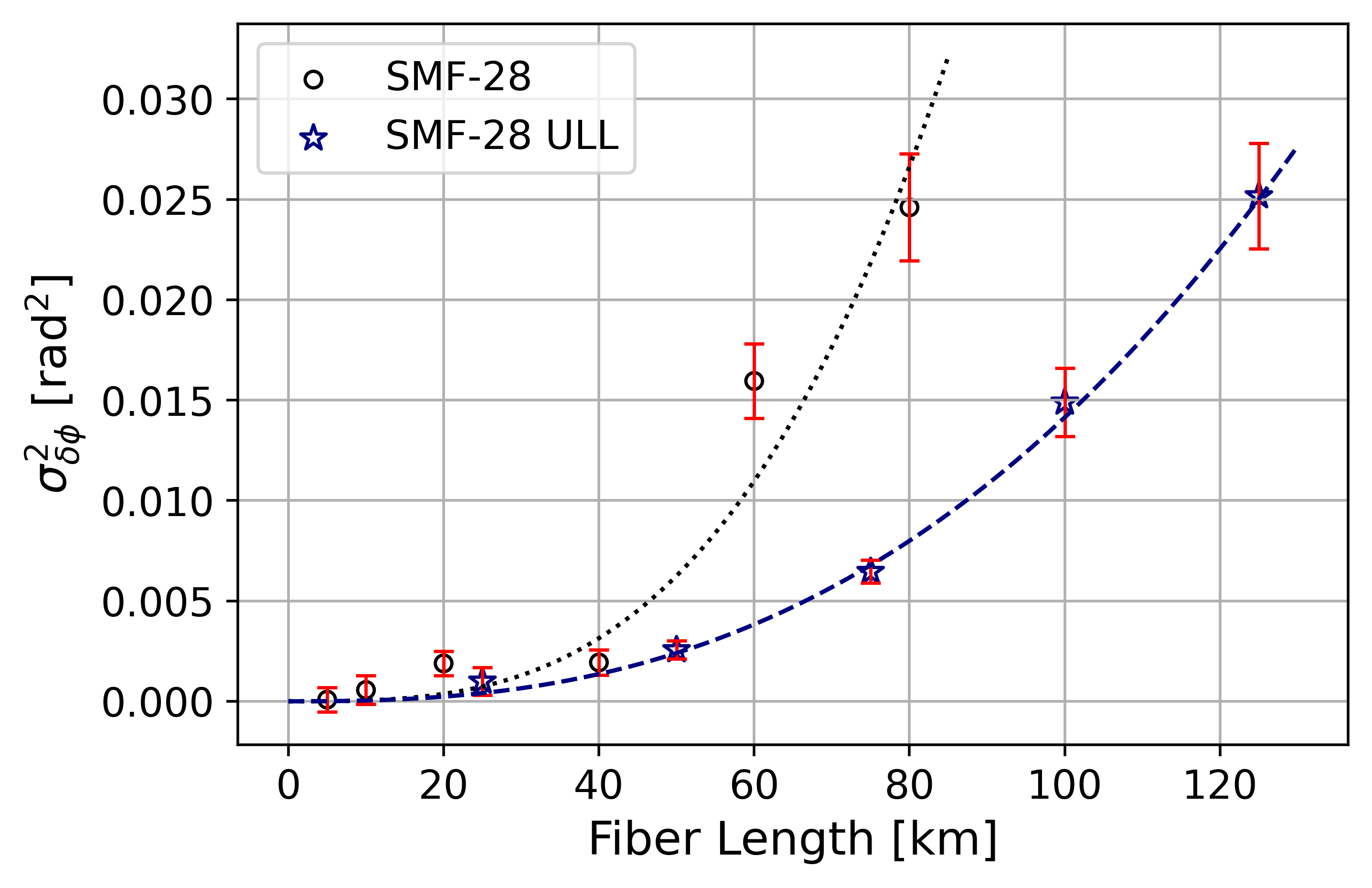} 
    \end{center}
    \caption{Variance of the fiber phase noise (calibrated) as a function of Sagnac loop length, $L$. Each data point is the average of ten data-sets, with the standard deviation used as the error bar. Curves represent a least-squares fit to $aL^b$, with $b=3.1(3)$ and $2.6(2)$ for SMF-28 and SMF-28 ULL fiber, respectively. } \label{fig:phi}
\end{figure}

We measure the Sagnac interference intensity fluctuations using the setup in Fig.~\ref{fig:sagnac}. Light from a CW distributed feedback laser diode (1545.3~nm) passes through a polarizer (not pictured) and is launched into the Sagnac loop. The PCs are adjusted such that the constructive interference signal, recorded by classical photodetector $D_0$, is maximized. This intensity signal is sent to an oscilloscope to record a time trace of the intensity, $I(\delta\phi(t))$. The purpose of analyzing the signal at an interference maximum is to ensure that the backscattering noise is not limiting the SNR in these measurements. 

$I(\delta\phi(t))$ was measured for loop lengths ranging from five to 125~km. Spools consisted of SMF-28 fiber (fiber attenuation coefficient, $\alpha=0.2$~dB/km) for all distances except 25~km, 50~km, 75~km, 100~km, and 125~km, which employed SMF-28 ULL fiber ($\alpha=0.16$~dB/km). A variable optical attenuator (VOA) is used to adjust the source intensity such that the average intensity at $D_0$ is the same for all loop lengths (roughly -25~dBm). We observe a significant correlation between the acoustic level in the laboratory and the magnitude of the intensity fluctuations. Therefore, effort was taken to minimize sounds in the lab while data was collected. Furthermore, to mitigate the effects of vibrations and thermal fluctuations in the environment, all fiber spools are stored inside a polyethylene box placed atop a pneumatically-isolated optical table. In a real-world Sagnac TFQKD network, fibers would likely be deployed underground, which may similarly assist in mitigating these noise effects as was observed in Ref.~\cite{Minar2008}. The intensity noise PSD was also recorded for the various loop lengths by replacing the oscilloscope with a RFSA. Details on these measurements are found in Appendix~\ref{app:phi_cal}.

From the measurement of $I(\delta\phi(t))$ and using Eq.~\ref{eq:I_int}, we can obtain the time-dependence of the phase fluctuation. A sampling rate of 100~MHz was used to record data over a total duration of 1~ms. These parameters ensure that we capture the phase fluctuations over the pulse period (100~ns) as well as over the transit time (0.02~ms to 0.61~ms) of the fiber. We evaluated the variance of the phase noise, $\sigma_{\delta\phi}^2$, by dividing the data-set composed of $N$ phase samples in time-ordered subsets of $n$ points. We then computed the variance of each subset and averaged over the number of subsets $N/n\approx10$.

Fig.~\ref{fig:phi} shows the results for $\sigma_{\delta\phi}^2$ as a function of Sagnac loop length, after equipment noise subtraction (details on this procedure are found in Appendix~\ref{app:phi_cal}). Each data point represents the average of ten data-sets and the error bar represents the standard deviation across the ten trials. We perform a least-squares fitting separately on the data from both types of fiber and find that $\sigma_{\delta\phi}^2$ increases as $L^{3.1(3)}$ and $L^{2.6(2)}$ in SMF-28 and SMF-28 ULL fiber, respectively. At first, the expectation arising from random walk theory is that the function plotted in Fig.~\ref{fig:phi} should scale as $L$. However, Clivati~\textit{et al.} showed that in Sagnac interferometry, where phase noise is passively compensated down to a limit imposed by the delay time between counterpropagating fields as they pass a given point in the fiber, the variance of the uncorrelated residual phase noise scales as $L^3$~\cite{Clivati2013}. Our measurements are shown to agree with this calculation. Interestingly, we observe that SMF-28 ULL fiber systematically outperforms SMF-28 fiber in terms of phase noise.

\section{Interference Visibility}\label{sec:visibility}
\begin{figure} [htbp]
    \centering \subfloat[\label{fig:iv_lengthdep}]{{\includegraphics[width=0.95\linewidth]{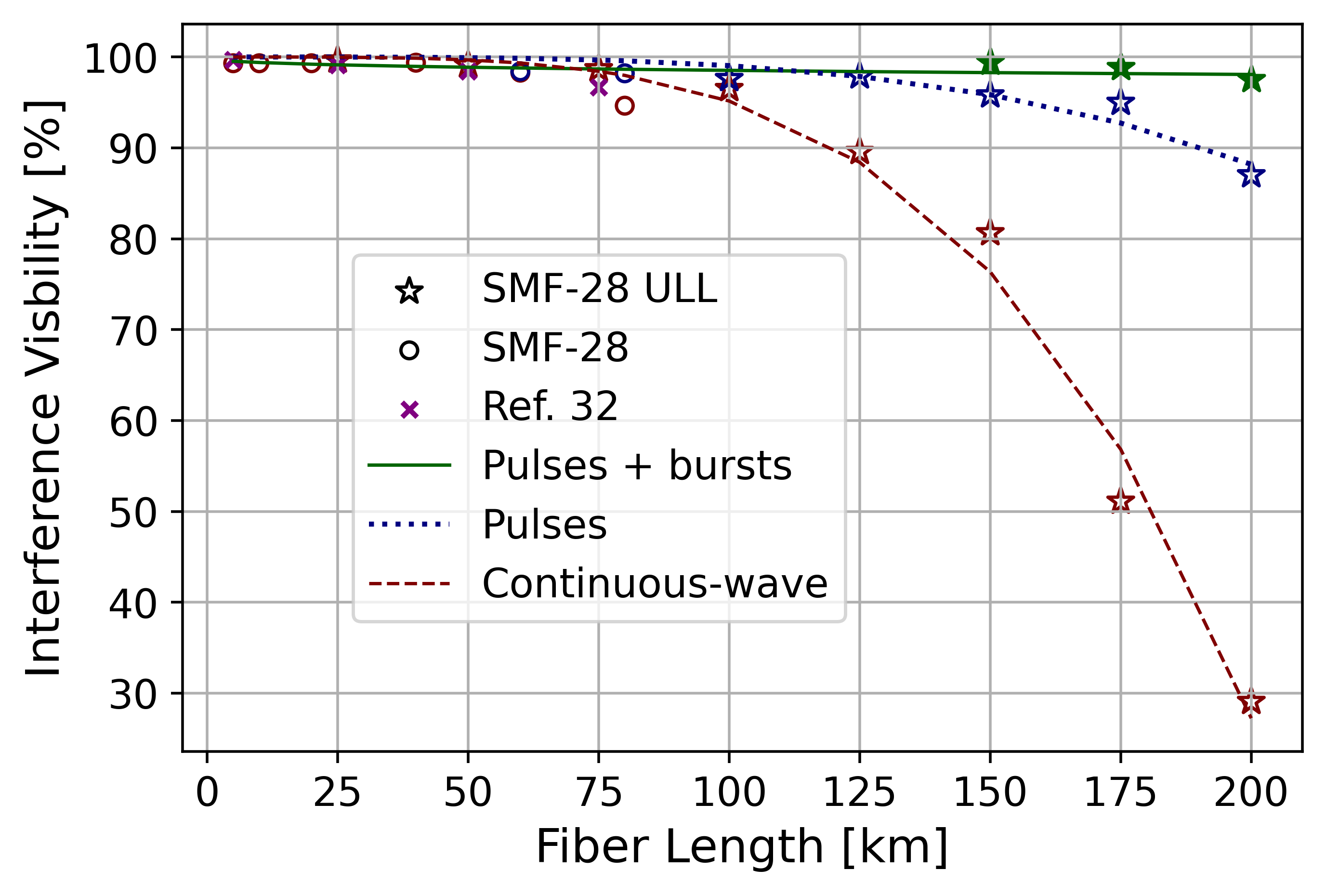}}}
    \qquad  
    \subfloat[\label{fig:interference}]{{\includegraphics[width=0.99\linewidth]{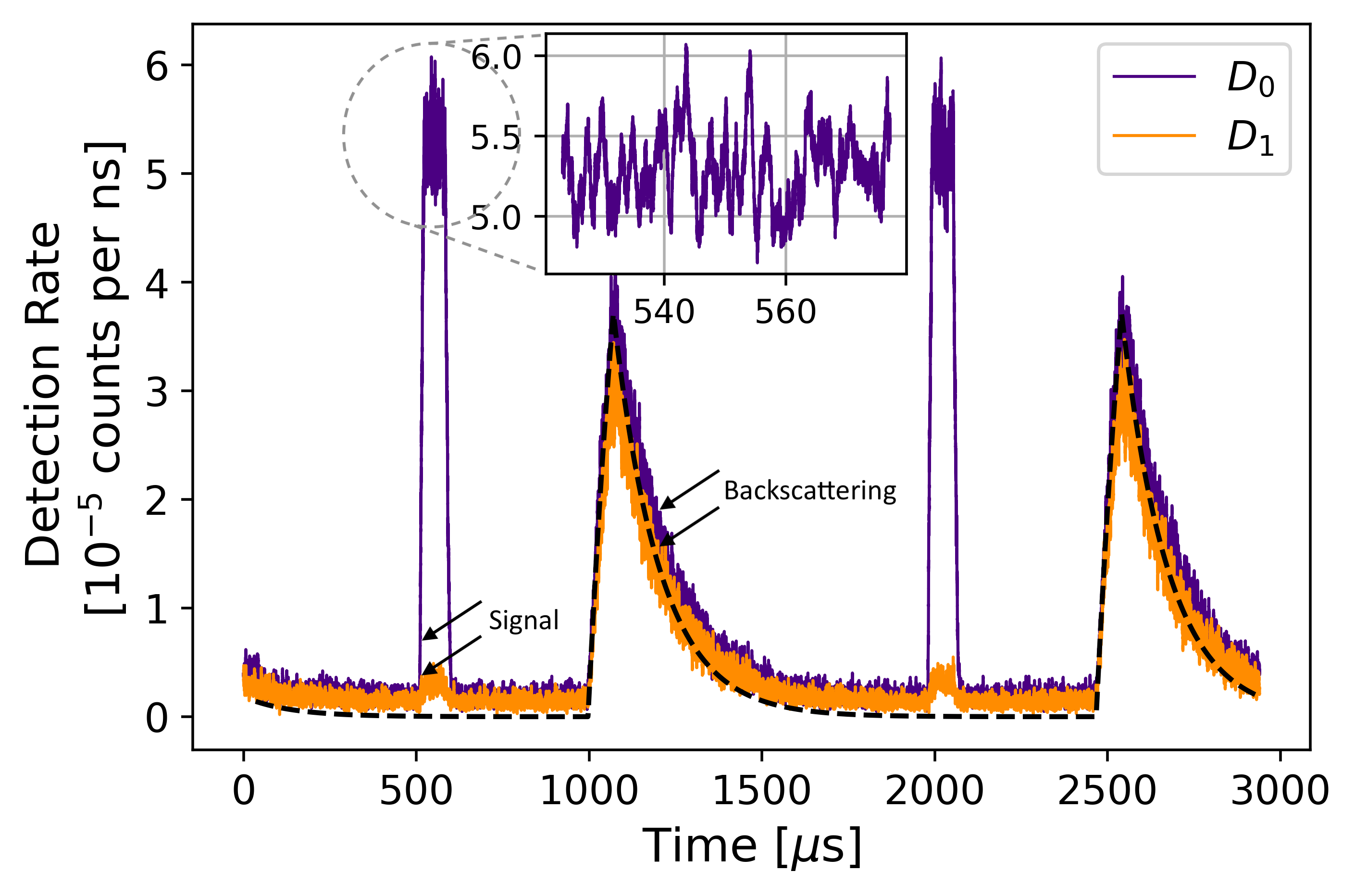}}}
    \qquad  
    \subfloat[\label{fig:pulse_window}]{{\includegraphics[width=0.99\linewidth]{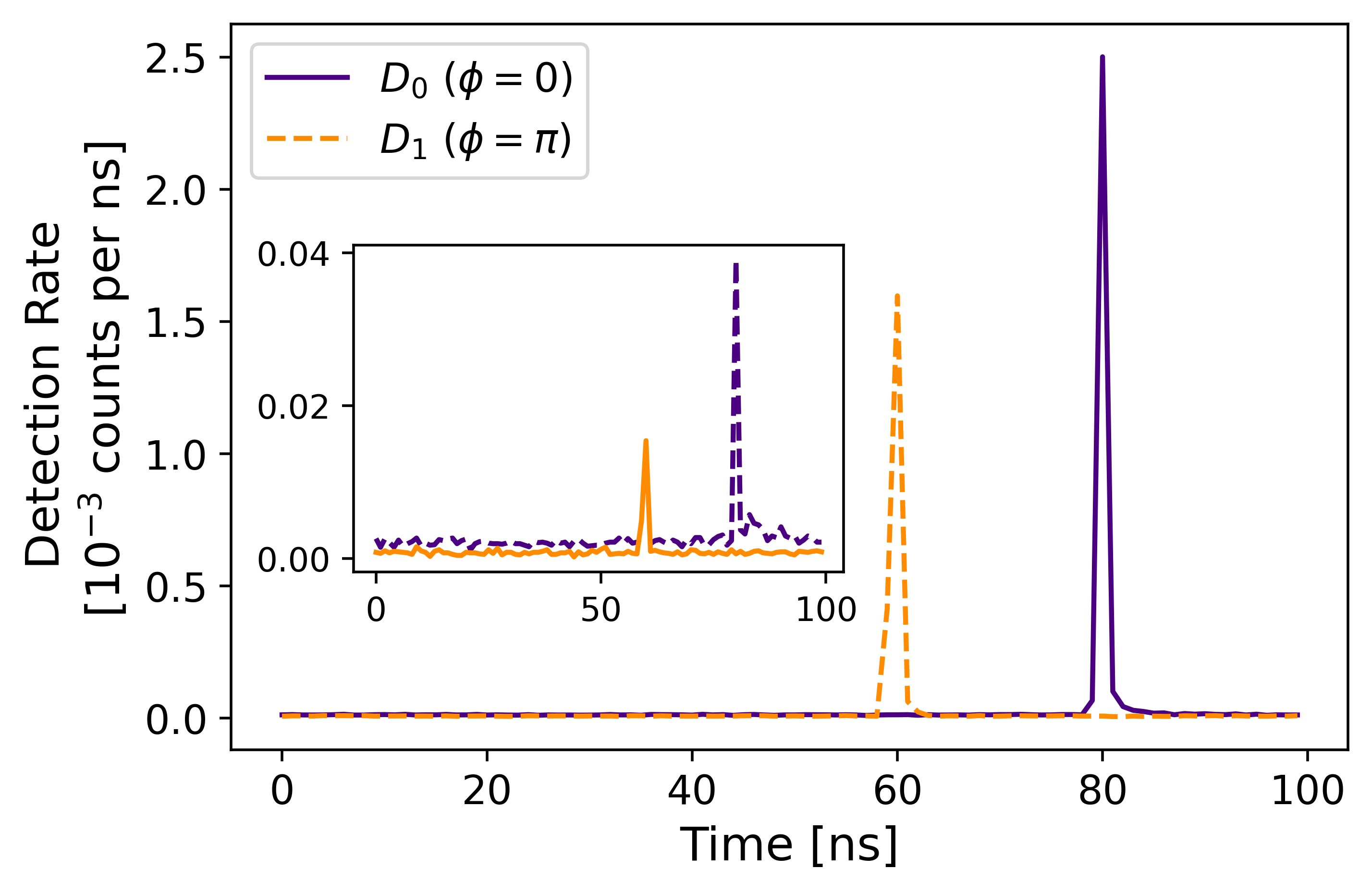}}}
    \caption{(a) Visibility (average of detectors $D_0$ and $D_1$) as a function of Sagnac loop length for different signal patterns. Curves serve as visual guides. (b) Counts registered over 30~s using single-photon detectors when burst-patterning is employed in 200~km fiber ($\phi=0$). A moving average over a 1~$\mu$s window is used. The backscattering noise, before it decays, is much stronger than the destructive interference signal ($D_1$). Inset shows intensity fluctuations due to fiber phase noise and Poisson noise. Dashed curve is the simulated backscattering noise (peak normalized to experimental data). (c) Raw counts within pulse detection windows.  
    }
    \label{fig:iv}
\end{figure}

We measure interference visibility using the setup in Fig.~\ref{fig:sagnac}. CW light passes through a 1x1 optical switch (extinction ratio $>25$~dB) driven by a square wave RF signal that determines when it is open or closed, thereby generating a burst pattern. An intensity modulator is used to generate pulses at a repetition rate of 10~MHz with 900~ps pulse width. Inside the Sagnac loop, a phase modulator with a polarizer is used to apply a $\phi=0$ or $\phi=\pi$ phase shift on only one of the interfering paths. A high-speed arbitrary waveform generator (Keysight M8195A) provides the synchronized RF signals to each of the modulators. 

The output of the interference is recorded by free-run SPADs (ID220) with an efficiency of 10\% and a dark count rate of about $7\times10^{-7}$ per pulse. Single-photon detectors must be used in order to detect the destructive interference signal for long fiber loops. A VOA is used to adjust the source intensity such that the overall detection rate (sum of both detectors) is the same for all loop lengths (roughly $3.5\times10^{-3}$ per pulse). In the absence of fiber phase noise, and if the interference minimum is set by the dark count rate, this detection rate would provide an optimal interference visibility of $99.96\%$. Measurements are recorded using an integration time of 30~s.   

We perform the experiment using three different signal patterns: CW light, 10~MHz pulse train, and 10~MHz pulse train sent in carefully timed on-off bursts with a 5\% duty cycle (details on the design of the burst pattern are presented in the Discussion). Note that when CW light is used, the phase modulator is removed from the loop and we go from maximum constructive interference in one detector to the other by adjusting the PCs. In each case, the visibility for each detector is calculated by Eq.~\ref{eq:visibility} and the average of the two is plotted in Fig.~\ref{fig:iv_lengthdep} as a function of fiber length. 

We achieve $>98\%$ visibility when using CW or pulsed light for fibers shorter than $80$~km, comparable to the results of Ref.~\cite{Minar2008}. The visibility at 200~km decreases to 29\% in the CW case and 87\% in the pulsed case, which is insufficient for TFQKD. The observed drop in visibility with longer fibers is due to two independent factors: backscattering noise and fiber phase noise. As we increase the fiber length, we must increase the source intensity in order to maintain a fixed overall detection rate, thereby increasing the backscattering noise. Eventually, the SNR becomes backscattering-limited, after which the visibility begins to decrease rapidly. Since backscattering is also proportional to the repetition rate of the signal, the visibility drops more rapidly when sending CW light than when sending a pulse train. If we instead send the pulses in bursts, we can mitigate the backscattering noise during the pulse detection windows and significantly improve the visibility, as was demonstrated in 150~km, 175~km, and 200~km fiber (filled data points in Fig.~\ref{fig:iv_lengthdep}). Ultimately, this technique allows us to assess the visibility that is limited by the fiber phase noise. 

To understand how this improvement arises from burst-patterning, consider Fig.~\ref{fig:interference} which shows the time-resolved detection rate for the 200~km Sagnac loop. We observe that the backscattering noise reaches levels that are much higher than the destructive interference signal. The use of bursts ensures that the backscattering has had time to sufficiently decay by the time we detect the pulses. The on-off time of the burst is different for each fiber length. For example, in the case of 200~km, pulses are on for 75~$\mu$s then off for 1400~$\mu$s. We use our simulation to ensure that the backscattering noise resulting from a given source intensity and signal pattern is suitable (dashed curve in Fig.~\ref{fig:interference}). Using this technique, we achieve a visibility of $97.4\%$ in a 200~km Sagnac loop (Fig.~\ref{fig:pulse_window}). We emphasize that this technique does not rely on any synchronization of the detectors with the RF signal generating the bursts. We recover the timing of the bursts using a Fourier method on the single-photon detections in post-processing, which allows us to resolve the pulse detection windows. 

\section{Discussion}\label{sec:discussion}
Our work demonstrates that in order to achieve high visibility in long-fiber Sagnac interferometers, a burst-patterning technique may be employed to accommodate backscattering noise. In Sagnac TFQKD, the fiber loop also contains an intensity modulator and a phase modulator for each QKD user~\cite{Zhong2019,Zhong2021,Zhong2022PRA}. One may show that if there are two users separated by a distance of $L/2$, we may select a burst on-time of $L/(2v_g)$ and an off-time of $L/v_g$, where $v_g$ is the group velocity of light in fiber, to prevent the backscattering from causing false detections. This burst pattern ensures that the intersection of clockwise and counterclockwise pulses takes place only in the region of the loop where the light from each direction has been attenuated by the loss of a user station. As a result, the intensity difference upon their crossing is low enough that the backscattered light which reaches the detectors at the same time as a signal photon is unlikely to generate a false count. With this configuration, the optimal burst duty cycle is given by~\cite{Xingye2023}
\begin{equation}\label{eq:burst_duty}
    d_B = \frac{L/(2v_g)}{L/(2v_g) + L/v_g} \approx 33\%.
\end{equation} This strategy of ensuring pulses intersect only after being attenuated is reminiscent of that of Ribordy~\textit{et al.} for plug \& play QKD~\cite{Ribordy2000}. However, our method does not rely on the use of an additional fiber storage line to set the burst on-time and our burst pattern can be optimized for an arbitrary Sagnac system by means of our backscattering simulation without changing or adding any hardware. 

In this work, since the loop does not contain the user stations, we still select a burst period of $3L/(2v_g)$ but fix $d_B=5\%$ to ensure that, for all fiber lengths, the backscattering reaches levels below the detector dark count during the pulse detection windows. Furthermore, we can expect that for a Sagnac system containing both long fibers and several lossy photonic devices, the advantage gained from burst-patterning will be even more significant than that observed in the current work. Consider that the SNR in a Sagnac interferometer (neglecting dark counts) scales as
\begin{equation}\label{eq:SNR}
    SNR = \frac{10^{-\beta/10}}{P_s(t)*x(t)},
\end{equation} where $\beta$ is the total loop loss in units of dB, $P_s(t)$ is the backscattering power returned from a single input pulse (proportional to the input pulse energy), and $x(t)$ is the input signal. For a configuration such as a GHz-speed Sagnac TFQKD network, $\beta$ is much larger than the loss due to fiber alone and $x(t)$ has a very high pulse repetition rate. Therefore, if no burst-patterning is employed, the visibility will become backscattering-limited at much shorter fiber lengths than observed in this work. 

The method of burst-patterning effectively reduces the efficiency of signal generation, in turn reducing the secure key rate. However, all other demonstrated TFQKD systems rely on measuring phase noise, which requires reserving a portion of the communication period for sending reference pulses that do not contribute to key generation. For example, the effective duty cycle for a 200~km channel in Ref.~\cite{Li2023} is $25\%$. Therefore, in terms of signal generation efficiency, Sagnac TFQKD is not disadvantaged compared to other configurations. Furthermore, Rayleigh scattering limits the SNR even in one-way TFQKD systems~\cite{Chen2020,Wang2022}, albeit at much longer transmission distances. This effect is due to double Rayleigh scattering of the strong reference pulses used for phase estimation. In such cases, the use of wavelength division multiplexing to allow for setting the wavelength of the reference light different from the quantum signal is needed to significantly reduce the double Rayleigh scattering noise~\cite{Pittaluga2021,Liu2023}.

Our findings help us to assess the fiber length limit of Sagnac TFQKD. Our measurement shows that we can achieve a phase-noise-limited interference visibility of $97\%$ in a 200~km loop, without active phase or polarization compensation. Based on Eq.~\ref{eq:v_vs_var}, this visibility corresponds to a phase noise variance $\sigma_{\delta\phi}^2 = 0.06$~rad$^2$, which agrees, within uncertainty, with the value extrapolated from the fit on our measurements in Fig.~\ref{fig:phi}. Ref.~\cite{Clivati2022} shows that in TFQKD we may evaluate the quantum bit error rate, $e$, by the relation $e=\sigma_{\delta\phi}^2/4$. Thus, our 200~km interference experiment would yield $e=2\%$, which is sufficient for secure key generation. We note that a 200~km fiber ring would support a communication distance of up to 100~km. Previously, we have also demonstrated a Sagnac TFQKD network where three users were placed along a fiber loop with 58~dB overall channel loss simulated with VOAs~\cite{Zhong2022PRA} (equivalent to 362~km SMF-28 ULL fiber). An asymptotic key rate of $R=2.014\times10^{-6}$ was obtained, comparable to the performance of other configurations (e.g., Ref.~\cite{Pittaluga2021} obtain $R=8.527\times10^{-7}$ over a 369~km channel). Therefore, while our system with auto phase stability is unlikely to achieve transmission distances as long as other setups for TFQKD, it can achieve similar key rates over equivalent channel losses, with the unique advantages of being low-cost and scalable to many users. This characterization suggests that such a system is feasible for settings such as ring metropolitan quantum networks~\cite{Herzog2004} which consist of long fibers as well as several lossy nodes. 

Lastly, we note that in Sagnac TFQKD where users receive and modulate light originating from an untrusted source, additional components should be implemented at each user station to guarantee security~\cite{Zhao2008,Zhao2010}. As discussed in Ref.~\cite{Zhong2019}, taps, photodetectors, and bandpass filters are necessary for Alice and Bob to detect and limit strong optical injections from the outside, so as to prevent eavesdroppers from probing the sources. A VOA is used to attenuate the pulses traveling back to Charlie to single-photon level. Thus, it is possible to achieve unconditional security with the Sagnac configuration for TFQKD while also maintaining inexpensive hardware requirements.

\section{Conclusion}\label{sec:conclusion}
In summary, we have characterized the visibility of long-fiber Sagnac interferometers up to a record 200~km in the presence of both phase and backscattering noise, and without active phase stabilization. We have shown that the backscattering noise problem can be solved with a burst-patterning technique to achieve a phase-noise-limited visibility of $>97\%$ in 200~km fiber. This work can also be extended to assess the scalability of several applications based on long-fiber Sagnac interferometers such as distributed sensing~\cite{Esmail2022}, quantum fingerprinting~\cite{Zhong2021QFP}, and quantum secret sharing~\cite{Bogdanski2009QSS}. 

\section*{Acknowledgements}
R.M. wishes to thank Xiaoqing Zhong, Andi Shahaj, Neel Choksi, Yen-An Shi, and Xingye Yang for helpful discussions. We also thank Corning for loaning some of the ULL spools. This work is supported by funding from NSERC, CFI, ORF, and MITACS.

\appendix
\section{Noise simulation and measurement of backscattering coefficient}\label{app:eta}
In this Appendix, we describe our model for simulating the backscattering noise as well as the procedure for measuring the backscattering coefficient. A fraction of light traveling through fiber will undergo Rayleigh scattering by small-scale inhomogeneities of the fiber permittivity, which act as an induced dipole moment. Indeed, this scattering process is what sets the lower limit to the fiber attenuation rate. Consider that the backscattering power $P_s(t)$ returned to the input end of a fiber from an input pulse of energy $P_0 \tau$ is~\cite{Brinkmeyer1980}
\begin{equation}\label{eq:P_s}
    P_s(t) = P_0\tau \eta e^{-\alpha v_g t},
\end{equation} where $\alpha$ is the fiber attenuation coefficient, $v_g$ is the speed of light in the fiber, and $\eta$ is the backscattering coefficient, which is the ratio of backscattering power (at t = 0) to forward pulse energy. We construct this backscattering response in fiber using an optical time-domain reflectometer (OTDR) setup. 

\begin{figure} [htbp] 
    \centering \subfloat[\label{fig:eta_setup}]
    {{\includegraphics[width=0.99\linewidth]{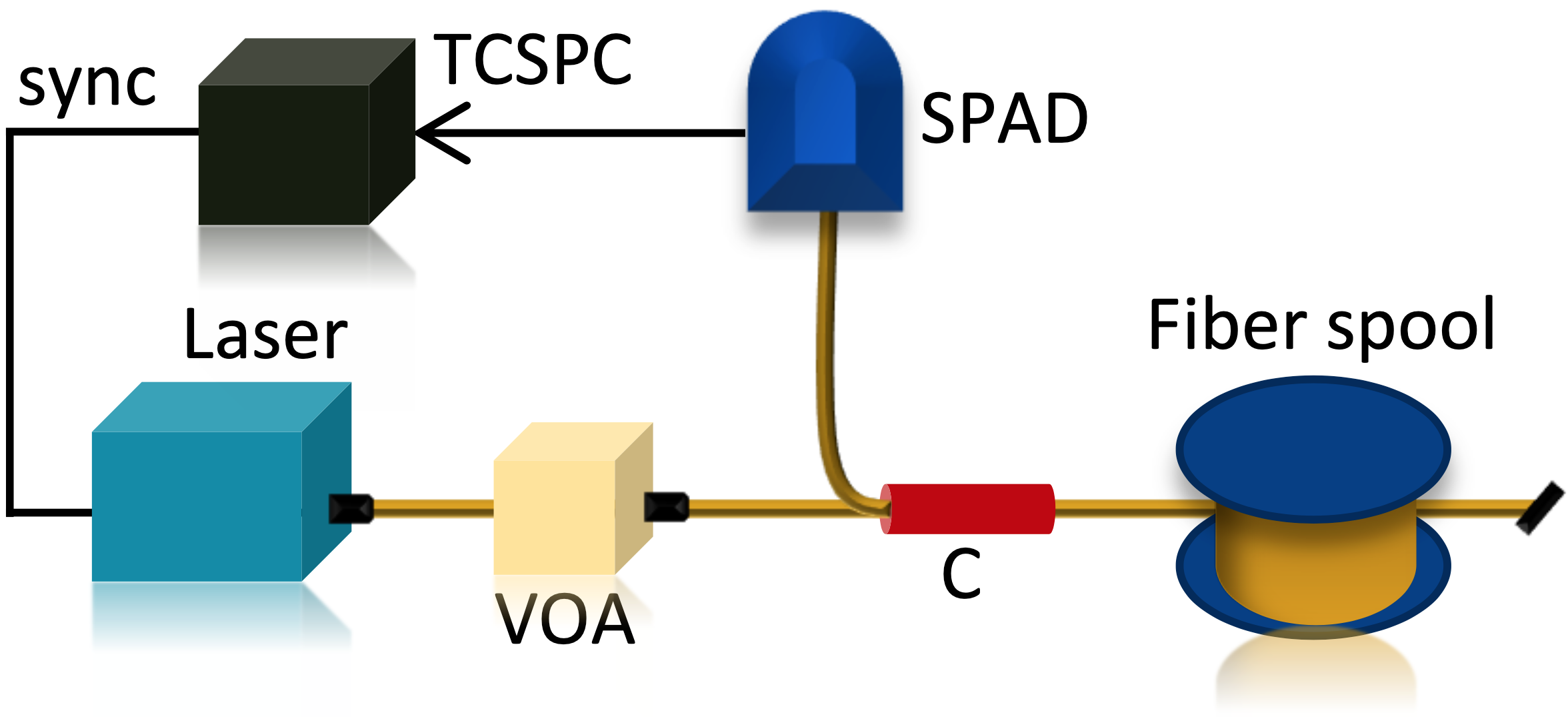}}}
    \qquad  
    \subfloat[\label{fig:eta_histogram}]{{\includegraphics[width=0.99\linewidth]{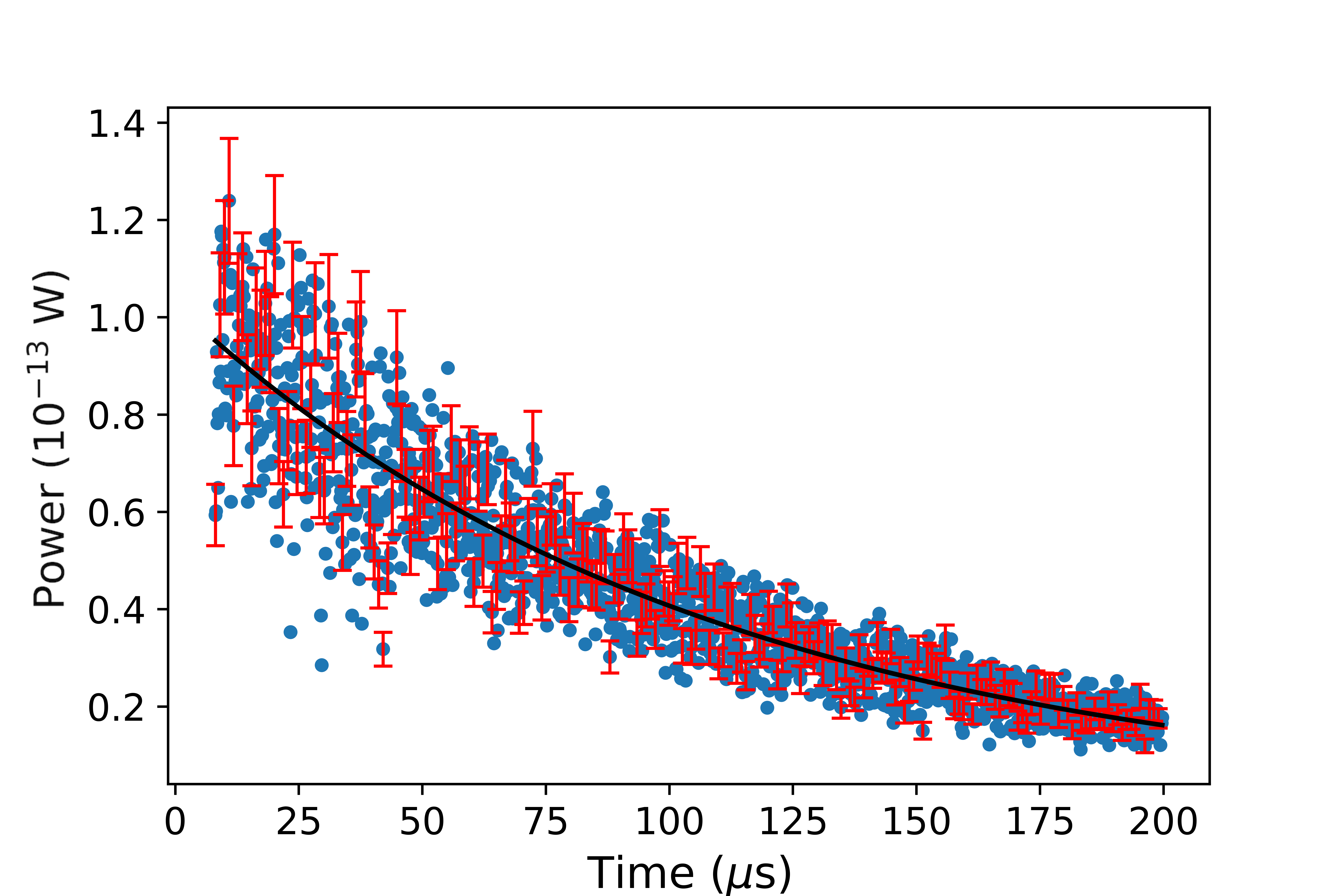}}}
    \caption{(a) Setup used to measure the backscattering coefficient, $\eta$. VOA: variable optical attenuator; C: circulator; SPAD: single-photon avalanche detector; TCSPC: time-correlated single photon counting system. (b) Backscattering response of 20~km SMF-28 fiber for an average input power of -72~dBm (6.5$\times$10$^{-11}$~W). Data is corrected for the dark count, dead-time, and detection efficiency of the SPAD. Error bars correspond to photon number uncertainty and dark count fluctuation and are plotted on a fraction of data points for visual clarity. Black curve is a least-squares fit to Eq.~\ref{eq:P_s}.
    }
    \label{fig:eta}
\end{figure}

Fig.~\ref{fig:eta_setup} depicts the OTDR experimental setup. A pulsed diode laser is used to generate short pulses (450~ps FWHM) at 5~kHz repetition rate, which are launched into 20~km SMF-28 fiber. The backscattered light is directed to a SPAD via a circulator. The repetition rate is selected such that the time between pulses is roughly the round-trip transit time in the fiber. Measurements are recorded using a time-correlated single photon counting system to construct the backscattering response over the signal period, as shown in Fig.~\ref{fig:eta_histogram}. A fitting of the data to Eq.~\ref{eq:P_s} for several different launch powers yields $\alpha$ = 0.202 $\pm$ 0.001~dB/km, which agrees with the fiber specification of 0.19~dB/km at 1550~nm, and $\eta$ = 8.0 $\pm$ 0.1~s$^{-1}$. An analogous procedure was used to measure $\eta$ in SMF-28 ULL fiber. The 20~km spool was replaced with a 50~km spool, and the pulse repetition rate was set to 2~kHz. Our measurements yield $\alpha$ = 0.159 $\pm$ 0.003~dB/km, which agrees with the fiber specification of 0.159~dB/km at 1550~nm, and $\eta$ = 6.54 $\pm$ 0.08~s$^{-1}$.

For reasonable pulse energies, Rayleigh backscattering in fiber can be modeled as a linear time-invariant system. Provided our pulses are sufficiently short compared to the time between the pulses, Eq.~\ref{eq:P_s} gives the system's impulse response. Therefore, to determine the backscattering from an arbitrary optical signal (e.g., a pulse train), we perform a convolution between Eq.~\ref{eq:P_s} and the signal~\cite{Phillips}. In order to account for a loop structure in the fiber, we consider the backscattering response of the clockwise and counterclockwise light independently. Note that the polarization of scattered light is random with respect to the incident light polarization. Therefore, the backscattering noise will on average be evenly distributed between the two detectors in a Sagnac interferometer. In the case that there are inserted components along the fiber, the impulse response of Eq.~\ref{eq:P_s} is modified to include discrete loss points at the appropriate locations. 

\section{Calibration of fiber phase noise and power spectral density measurements}\label{app:phi_cal}
\begin{figure} [H]  
    \centering \subfloat[\label{fig:var}]
    {{\includegraphics[width=0.99\linewidth]{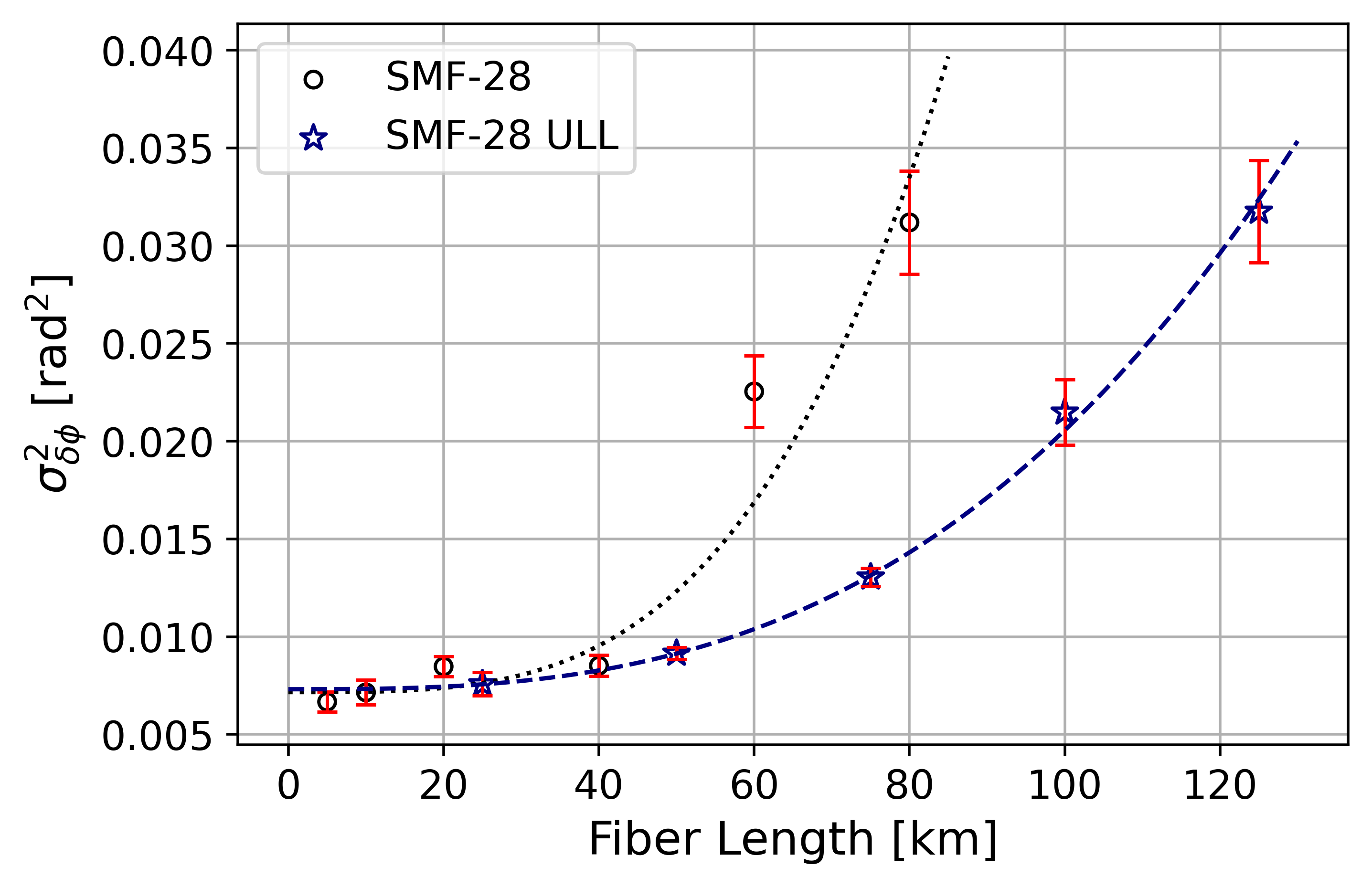}}}
    \qquad  
    \subfloat[\label{fig:psd}]{{\includegraphics[width=0.99\linewidth]{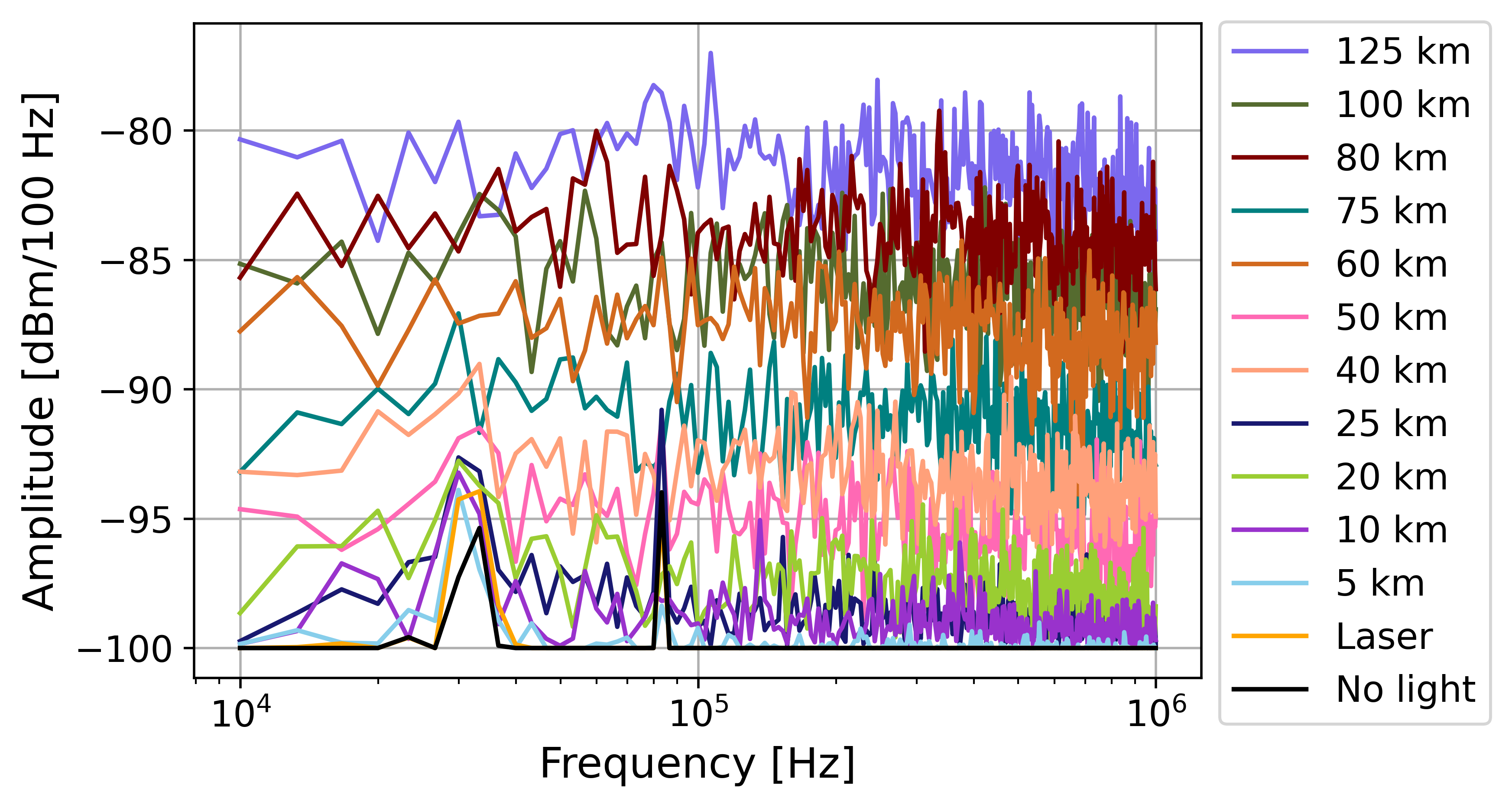}}}
    \caption{(a) Variance of the phase noise (pre-calibration) as a function of Sagnac loop length, $L$. Each data point is the average of ten data-sets, with the standard deviation used as the error bar. Curves represent a least-squares fit to $aL^b+c$, with $c=0.0072(3)$~rad$^2$ and $0.0073(6)$~rad$^2$ for SMF-28 and SMF-28 ULL fiber, respectively. (b) Intensity noise power spectral density (PSD) for various Sagnac loop lengths. Frequencies are scanned between 9~kHz and 1~MHz with a resolution bandwidth of 100~Hz. Each trace shown is the average of ten data-sets. A PSD of the laser output and the electronic noise are included for reference.
    }
    \label{fig:phi_precal}
\end{figure}

In this Appendix, we detail the procedure for isolating the fiber phase noise from the equipment (photodetector) noise and present measurements on the intensity noise PSD of Sagnac interference. Fig.~\ref{fig:var} shows the results for the variance of the phase noise, $\sigma_{\delta\phi}^2$, as a function of Sagnac loop length, before equipment noise subtraction. We see that for short fibers, the phase noise approaches a constant non-zero value. A least-squares fitting to $aL^b + c$, separately on the data from both types of fiber, returns an average y-intercept of $c=0.0072(3)$~rad$^2$. Moreover, from the PSD in Fig.~\ref{fig:psd}, we observe that the noise in short fibers ($<40$~km) is comparable to the level of noise in the detector. Therefore, we subtract the lower bound of $c$ from $\sigma_{\delta\phi}^2$ in order to obtain the noise due to fiber shown in Fig.~\ref{fig:phi}.

Fig.~\ref{fig:psd} shows the PSD measurements, obtained using the same procedure outlined in Sec.~\ref{sec:psd}, with a RFSA replacing the oscilloscope. Measurements were taken with a resolution bandwidth of 100~Hz. Each trace has a frequency range between 9~kHz and 1~MHz and is the average of ten data-sets collected consecutively over the span of about twenty minutes. The frequency range is chosen to capture intensity fluctuations over the pulse period as well as over the transit time of the fiber (within the RFSA limitations). The standard deviation of the ten measurements (not pictured) is 1~dB for five km, and increases to 4~dB for 125~km. We also measured the PSD of the laser output (after the polarizer) to show that the laser intensity fluctuations are negligible compared to the electronic noise. We observe spikes in the PSD at roughly 20~kHz and 47~kHz, which could be due to electromagnetic interference. For each loop length, the intensity noise is observed to fluctuate randomly about a constant value over all the scanned frequencies. The integrated noise level is found to increase with loop length in a manner similar to $\sigma_{\delta\phi}^2$.

\bibliography{refs}

\end{document}